\documentclass[journal=jacsat,manuscript=article]{achemso}

\usepackage{chemformula} 
\usepackage[T1]{fontenc} 
\usepackage{bm}
\usepackage{mathrsfs}
\usepackage{amssymb}

\newcommand{\ex}{\mathrm{e}}

\newcommand{\bd}{\bm{D}}
\newcommand{\br}{\bm{r}}
\newcommand{\bk}{\bm{k}}
\newcommand{\eh}{\hat{e}}
\newcommand{\be}{\bm{\mathcal{E}}}

\author{Mohammad Reza Karimpour}
\affiliation{Department of Physics and Materials Science, University of Luxembourg, L-1511 Luxembourg City, Luxembourg}
\author{Dmitry V. Fedorov}
\affiliation{Department of Physics and Materials Science, University of Luxembourg, L-1511 Luxembourg City, Luxembourg}
\author{Alexandre Tkatchenko}
\affiliation{Department of Physics and Materials Science, University of Luxembourg, L-1511 Luxembourg City, Luxembourg}
\email{alexandre.tkatchenko@uni.lu}

\title[An \textsf{achemso} demo]
  {Molecular Interactions Induced by a Static Electric Field in Quantum Mechanics and Quantum Electrodynamics}

\abbreviations{IR,NMR,UV}

\begin{document}

%
%
%
%
%

\newpage
\begin{abstract}
By means of quantum mechanics and quantum electrodynamics applied to coupled harmonic Drude oscillators, we study the interaction between two neutral atoms or molecules subject to a uniform static electric field. Our focus is to understand the interplay between leading contributions to field-induced electrostatics/polarization and dispersion interactions, as considered within the employed Drude model for both nonretarded and retarded regimes. For the first case, we present an exact solution for two coupled oscillators obtained by diagonalizing the corresponding quantum-mechanical Hamiltonian and demonstrate that the external field can control the strength of different intermolecular interactions and relative orientations of the molecules. In the retarded regime described by quantum electrodynamics, our analysis shows that field-induced electrostatic and polarization energies remain unchanged (in isotropic and homogeneous vacuum) compared to the nonretarded case. For interacting species modeled by quantum Drude oscillators, the developed framework based on quantum mechanics and quantum electrodynamics yields the leading contributions to molecular interactions under the combined action of external and vacuum fields.
\end{abstract}

\begin{figure}[h!]
\includegraphics[width=0.48\linewidth]{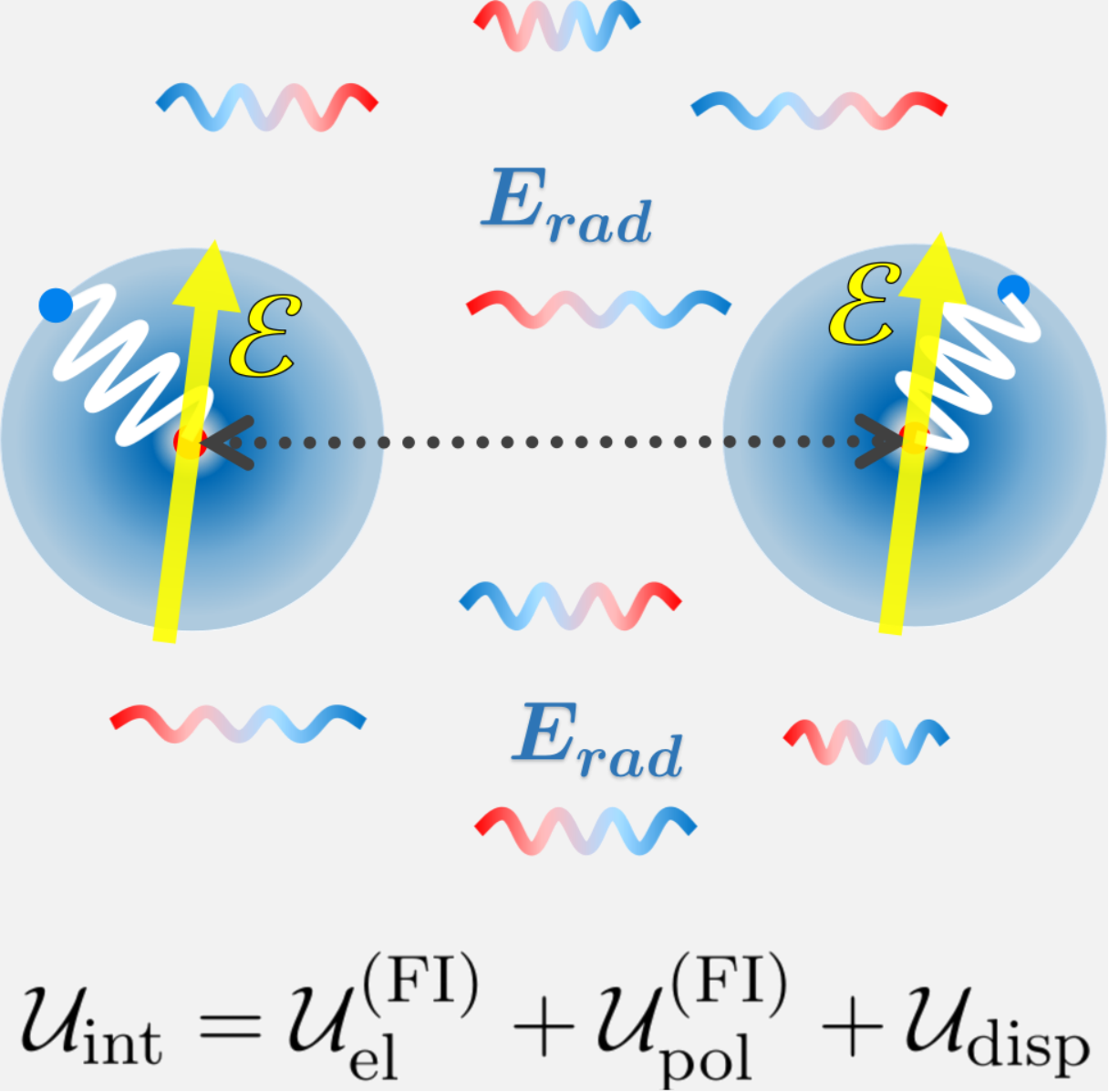}
\end{figure}

\newpage
The stable structure and properties of biomolecules, nanostructured materials, and molecular solids are determined by a delicate balance between different intermolecular forces~\cite{Kaplan2006,Stone2013,Tkatchenko2015,Langbein_1974,Dobson_book_2020}. In many realistic systems, molecular interactions are substantially modified by solvents, cell membranes, ionic channels, and other environments~\cite{Sadhukhan2017,Stoehr2021,Stoehr2019-SciAdv,Kleshchonok2018}. A proper description of such environments demands robust approaches for modeling both nonretarded and retarded intermolecular interactions under arbitrary fields.
Molecular interactions in presence of static and dynamic electromagnetic and thermal fields have been studied using various approaches~\cite{Silberstein_1917_1, Silberstein_1917_2, Buckingham_1956, Buckingham_1973, Buckingham_1978, Hunt1985, Hunt1986, Thirunamachandran1980, Rubio2021,Milonni1996,Sukhov2013,Brugger2015,Marinescu1998,Muruganathan2015,Kleshchonok2018,Fiscelli2020}, but a comprehensive understanding is still missing and some results remain controversial. For example, random and inhomogeneous fields have been shown to affect the strength and distance dependence of van der Waals (vdW) interactions or even change their sign~\cite{Thirunamachandran1980,Rubio2021,Milonni1996,Sukhov2013,Brugger2015,Sadhukhan2017}. 
However, similar to the discussion on the textbook dispersion interaction~\cite{Feynman1939,Hunt_1990}, there is still an ongoing debate on the interpretation of modified vdW interactions as having either an electrostatic
or a quantum-mechanical origin~\cite{Sadhukhan2017,comment-to-Mainak,Mainak-reply,Stoehr2021}. 
The application of weak static (in)homogeneous fields in the nonretarded regime~\cite{Marinescu1998,Muruganathan2015,Kleshchonok2018} yields a visible modification of molecular interactions in second and third orders of perturbation theory, however the retarded regime has not been addressed in these studies. Fiscelli \textit{et al.}~\cite{Fiscelli2020} used quantum electrodynamics (QED) to propose a \emph{dispersion} energy, scaling as $R^{-4}~(R^{-3})$, with respect to the interatomic distance for retarded (nonretarded) regime, for interacting two-level hydrogen-like atoms in static electric fields. However, it still remains unclear~\cite{Comment_on_Fiscelli2020, Fiscelli_reply} whether this term corresponds to dispersion or electrostatic interactions. In order to resolve existing controversies and clarify discrepancies in the literature, here we develop
a quantum framework for modeling and understanding intermolecular
interactions in electric fields based on first principles of quantum mechanics and QED.  

The advances made in this Letter hinge on the usage of two formalisms that enable an accurate modeling and conceptual understanding of nonretarded and retarded interactions for two coupled quantum Drude oscillators (QDO)~\cite{Wang2001, Sommerfeld2005, Jones2013} subject to a static 
electric field: solving Schr\"odinger's equation via exact
diagonalization and using perturbation theory in QED~\cite{Cohen-Tannoudji1997,Milonni1994,Craig1994,Salam2009,Buhmann2012,Passante2018,Rubio2018}. In addition, we employ stochastic electrodynamics (SED)~\cite{Marshall1963,Marshall1965,Boyer1975,Pena1978,Boyer1980,Pena1996,Nieuwenhuizen2019,Boyer2019}, as a semiclassical formalism that transparently connects molecular interactions to the fields that originate them.
The usage of QDOs to accurately and efficiently model the linear response of valence electrons in atoms and molecules is a critical aspect because coupled QDOs enable analytical solutions (with and without electric field) and have been convincingly demonstrated to provide a reliable quantitative tool to describe response properties of real atoms and molecules subject to external fields or confinement~\cite{Wang2001,Sommerfeld2005,Jones2013,Tkatchenko2012,Reilly2015,Gobre2016,Sadhukhan2016,Hermann2017,Fedorov2018,Tkatchenko2021}. QDOs can quantitatively -- within a few percent compared to explicit treatment of electrons -- describe polarization and dispersion interactions~\cite{Jones2013,Sadhukhan2016,Hermann2017}, capture electron density redistribution induced by these interactions~\cite{Hermann-NatureComm}, model intermolecular interactions in electric fields~\cite{Sadhukhan2017,Kleshchonok2018}, among many other response phenomena~\cite{Tkatchenko2015}. Our current study benefits from many attractive features of QDOs and demonstrates their applicability to the retarded regime.  
By means of the developed framework, we derive dominant contributions to the interaction energy of two QDOs in an electric field up to terms $\propto R^{-6}\,(R^{-7})$ for nonretarded (retarded) regime. 
These contributions, corresponding to a linear response to the external field, are interpreted as the field-induced electrostatic and polarization 
interactions obtained in addition to the conventional leading-order London/Casimir dispersion interaction found to be unchanged in the presence of a static electric field within the QDO model.
We note that most previous studies in the QED literature employed the two-level ($s$ and $p$ states) hydrogen-like atom as a model. Unfortunately, there is no known analytical solution for the case of interacting hydrogen atoms under an external electric field~\cite{Fiscelli2020}, causing quite some controversy over the interpretation of field-induced interatomic interactions.~\cite{Comment_on_Fiscelli2020, Fiscelli_reply, Hu2021} Of course, the coupled QDO model employed here also introduces approximations. In particular, the Gaussian form of the QDO wavefunction does not capture the effect of deformation of an electron cloud by a static field, in contrast to a hydrogen atom for example. In a homogeneous electric field, the ground-state electron density of a QDO undergoes a rigid displacement and this means that the $\beta$ and $\gamma$ hyperpolarizabilities vanish~\cite{Jones2013}. Therefore, field-induced and dispersion-induced changes in the polarizability of interacting species described by these hyperpolarizabilities \cite{Jansen1955, Buckingham_1956, Buckingham_1973, Buckingham_1978, Hunt1980, Hunt1986, Hunt_1994, Fowler1994, Certain1971, Joslin_1996} are missing within the QDO model~\cite{Hunt1981, Jansen1955}. Hence, some interactions corresponding to the coupling of the field-induced and dispersion-induced changes in polarizabilities 
will also vanish for coupled QDOs. However, as we discuss below, such terms are either of higher order ($\beta$-terms) or smaller magnitude ($\gamma$-terms) in comparison to the 
dispersion and field-induced electrostatic energies, for weak static electric fields.

\begin{figure}[ht!]
\includegraphics[width=0.8\linewidth]{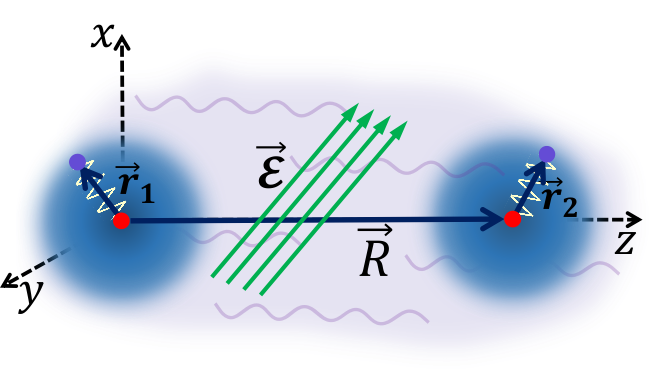}
\caption{Two interacting quantum Drude oscillators under influence of both, the fluctuating vacuum electromagnetic field and an applied uniform
static electric field ${\bm{\mathcal{E}}} = (\mathcal{E}_x,\mathcal{E}_y,\mathcal{E}_z)$.}
\label{fig:QDOs}
\end{figure}

First, we consider the nonretarded case, when the distance $R$ between two interacting species is small compared to wavelengths $\lambda_e$ of their electronic transitions: $R \ll\lambda_e =  2\pi c/\omega_e\,$, where $c$ is the speed of light. 
The system of two interacting
QDOs, which represent two atoms or molecules, separated by a distance $R$ along the $z$ axis is shown
in Fig.~\ref{fig:QDOs}. In the absence of any field, this system is described by the Hamiltonian
\begin{align}
\!\!\!
H\! =\! -\tfrac{\hbar^2}{2} \!\left[\!\tfrac{\boldsymbol\nabla_{\bm{r}_1}^2}{m_1}
\!+\! \tfrac{\boldsymbol\nabla_{\bm{r}_2}^2}{m_2} \!\right]\! +\! \tfrac{m_1\omega_1^2 r_1^2+ m_2\omega_2^2 r_2^2}{2} + 
\!V_{\rm int}(\bm{r}_1, \bm{r}_2)\,,\!\!\!\!
\label{nret-Hamiltonian-1}
\end{align}
where $V_{\rm int}$ is the Coulomb potential approximated here by
the dipole-dipole coupling
\begin{equation}
\!\!
V_{\rm int}(\bm{r}_1, \bm{r}_2) 
\approx V_{\rm int}^{\rm dd}(\bm{r}_1, \bm{r}_2) = \tfrac{q_1 q_2}{4\pi\epsilon_0 R^3} \left(\bm{r}_1\cdot\bm{r}_2 - 3z_1 z_2 \right) . \!\!\!
\label{nret-V}
\end{equation}
This form of the interaction breaks the initial symmetry between $z$ and $x/y$ parts of the Hamiltonian.
However, the symmetry between $x$ and $y$ parts remains. Thus, one needs to obtain just $H_x$ and $H_z$ from the separation $H = H_x + H_y + H_z$. Introducing the normal-mode coordinates
\begin{equation}
x_\pm =\pm\, \tfrac{\gamma_x \sqrt{m_1} x_1 + [(\omega_2^2 - \omega_1^2)/2 \pm \sqrt{D}] \sqrt{m_2} x_2}
{\sqrt{[(\omega_2^2 - \omega_1^2)/2 \pm \sqrt{D}]^2 + \gamma_x^2}}\ ,
\label{new_coordinates}
\end{equation}
with $\gamma_x = \frac{q_1 q_2}{4\pi\epsilon_0 \sqrt{m_1 m_2} R^3}$ and
$D = ((\omega_2^2 - \omega_1^2)/2)^2 + \gamma_x^2\,$, we transform the $x$-dependent part of Eq.~\ref{nret-Hamiltonian-1} to
\begin{equation}
H_x = -\tfrac{\hbar^2}{2}\left[\tfrac{d^2}{d x_+^2} + \tfrac{d^2}{d x_-^2}\right]
+ \tfrac{1}{2}(\omega_+ x_+^2 + \omega_- x_-^2)\ ,
\label{nret-Hamiltonian-x2}
\end{equation}
where $\omega_\pm^2 = \frac{(\omega_2^2 + \omega_1^2)}{2} \pm \sqrt{D}$. The Hamiltonian of Eq.~\ref{nret-Hamiltonian-x2} describes two independent QDOs with unit masses and
the frequencies $\omega_+$ and $\omega_-$ of the two normal modes.

Now we introduce a uniform electric field with the contribution
$H_{\rm f}=-(q_1\bm{r}_1+q_2\bm{r}_2)\cdot\bm{\mathcal{E}}$ to the Hamiltonian. With the new coordinates, we obtain
\begin{equation}
H'_x \equiv H_x + H_{\rm f} = H_x
- \mathcal{E}_x ( f_+ x_+ + f_- x_-)\ ,
\label{nret-Hamiltonian-x3}
\end{equation}
where $H_x$ is given by Eq.~\ref{nret-Hamiltonian-x2} and
\begin{equation}
f_{\pm}=\tfrac{(q_1/\sqrt{m_1})\gamma_x + (q_2/\sqrt{m_2}) [(\omega^2_2 - \omega^2_1)/2\pm \sqrt{D}]}{\sqrt{\gamma_x^2 + [(\omega^2_2 - \omega^2_1)/2 \pm \sqrt{D}]^2}} \ .
\end{equation}
Equation~\ref{nret-Hamiltonian-x3} can be rewritten as a quadratic form
\begin{equation}
H'_x = \textstyle\sum\limits_{i = \pm} \left[ -\tfrac{\hbar^2}{2} \tfrac{d^2}{d x_i^2} + \tfrac{1}{2} \omega_i^2
\left(x_i -\tfrac{f_i\mathcal{E}_x}{\omega_i^2}\right)^2
- \tfrac{f_i^2 \mathcal{E}_x^2}{2\omega_i^2} \right]\ ,
\label{nret-Hamiltonian-x4}
\end{equation}
which is the Hamiltonian of two one-dimensional (1D) oscillators with the frequencies $\omega_+$ and $\omega_-$ and the centers shifted by $(f_+\mathcal{E}_x/\omega_+^2)$ and
$(f_-\mathcal{E}_x/\omega_-^2)$, respectively. 
The interaction energy of the two oscillators under an external electric field is given by the difference between the total
energy of the coupled QDOs and the sum of the total energies of
two isolated QDOs in the same field
\begin{equation}
\label{nret-energy-in-f-x}
\Delta \mathcal{U}_{x} = \textstyle\sum\limits_{i = \pm} \left[ \tfrac{\hbar\omega_i}{2} - \tfrac{f_i^2 \mathcal{E}_x^2}{2\omega_i^2} \right] - \textstyle\sum\limits_{i = 1,2} \left[ \tfrac{\hbar\omega_i}{2} - \tfrac{\alpha_i^2 \mathcal{E}_x^2}{2} \right]\ ,
\end{equation}
where $\alpha_i = q_i^2/m_i\omega_i^2$ is the static polarizability of a QDO. Due to the symmetry, $\Delta \mathcal{U}_y$ is derived in the same way as $\Delta \mathcal{U}_x$ by replacing subscripts $x$ with $y$. 
The $z$-dependent part of the total Hamiltonian, $\Delta \mathcal{U}_z$, one can obtain similarly to $\Delta \mathcal{U}_x$, by replacing $\gamma_x$ in Eq.~\ref{new_coordinates} with $\gamma_z=-2\gamma_x$.
For this case, we obtain 
$z_\pm =\left(\gamma_z z'_1 + [(a_2 - a_1) \pm \sqrt{D_z}] z'_2\right)/\sqrt{\gamma_z^2 + [(a_2 - a_1) \pm \sqrt{D_z}]^2}$ and $\omega_\pm = [(a_1 + a_2) \mp \sqrt{D_z}]^{\nicefrac 12}$, where $D_z = (a_2 - a_1)^2 + \gamma_z^2$.

As shown in Ref.~\citenum{Our_arXiv}, the resulting expressions for the interaction energy can be expanded as infinite series with respect to small terms proportional to $\alpha_1\alpha_2/(4\pi\epsilon_0)^2 R^6$. Retaining all the leading terms up to $R^{-6}$ within the QDO model, we obtain 
\begin{align}
\label{nret-energy}
\Delta\mathcal{U}(R) = -\tfrac{3~\alpha_1 \alpha_2 \hbar \omega_1 \omega_2}{2 (4\pi\epsilon_0)^2 (\omega_1+\omega_2) R^6}
+ \tfrac{\alpha_1 \alpha_2 (\mathcal{E}_x^2+\mathcal{E}_y^2-2\mathcal{E}_z^2)}{4\pi \epsilon_0 R^3} - \tfrac{\alpha_1 \alpha_2 (\alpha_1+\alpha_2) (\mathcal{E}_x^2+\mathcal{E}_y^2+4\mathcal{E}_z^2)}{2 (4\pi\epsilon_0)^2 R^6}\ .
\end{align}
Here, the first contribution is the well-known vdW dispersion energy, stemming from the difference between $\hbar\omega_i/2$ terms in Eq.~\ref{nret-energy-in-f-x} as well as the corresponding expressions for $\Delta \mathcal{U}_y$ and $\Delta \mathcal{U}_z$~\cite{Our_arXiv}. The dispersion energy is not affected by the static field within the QDO model. The other two terms in Eq.~\ref{nret-energy} are field-induced contributions, which originate from the terms in Eq.~\ref{nret-energy-in-f-x} not containing $\hbar$. As will be discussed in detail below, these terms correspond to the field-induced electrostatic and polarization contributions to the interaction energy, respectively. For real atoms or molecules,  Eq.~\ref{nret-energy} would contain an additional term $\propto R^{-6}$, which has the approximate form~\cite{Buckingham_1956,Buckingham_1978,Joslin_1996} (for an exact analytical expression, see Ref.~\citenum{Hunt1986}) $\Delta\mathcal{U}_{\gamma} (R) = - \gamma C_6 (4 \mathcal{E}_x^2 + 4 \mathcal{E}_y^2 + 7 \mathcal{E}_z^2)/18 \alpha R^6$, where $\gamma$ is the second hyperpolarizability and $C_6$ is the dipole-dipole vdW dispersion coefficient. This contribution is absent within the QDO model where $\gamma$ vanishes because of symmetry reasons~\cite{Jansen1955,Hunt1981}. In addition, it is worth mentioning that vdW dispersion interaction between two atoms or molecules also causes dispersion-induced dipoles~\cite{Feynman1939, Hunt_1980_dipole, Galatry1980, Galatry1983, Hunt1985, Craig_1981, Hunt_1987, Hunt_1987_2, Fowler_1990, Vigoureux1983} that vary as $R^{-7}$. With coupling to an applied static field, it results in an interaction energy which is linear in the field and scales as $R^{-7}$. 
As first suggested by Hunt~\cite{Hunt_1980_dipole}, such an interaction energy corresponds to the hyperpolarization of an atom by the fluctuating field from the neighboring 
atom in combination with the applied field, which depends on the $B$ hyperpolarizability (dipole-dipole-quadrupole hyperpolarizability) and the static dipole polarizability of the atoms~\cite{Hunt_1980_dipole}. 
Since a QDO has a nonvanishing $B$ hyperpolarizability, the dispersion-induced dipole moment can be captured by taking into account dipole-quadrupole couplings \cite{Odbadrakh2016} in the interaction Hamiltonian of Eq.~\ref{nret-V}.

In order to extend our results to the retarded regime, $\!R\!\gg\!c/\!\omega_e$, we consider the coupling between two QDOs in the framework of QED, where the molecular interactions are mediated by the fluctuating vacuum radiation field. 
The total system consists of two QDOs, an external static field, and the vacuum field. 
For a single QDO coupled to the static field, the Hamiltonian is given by 
\begin{equation}
\label{nonperturbed-HQDO}
H^{(\mathcal{E})} \equiv H^{(\mathcal{E})} (\bm{r}) 
= - \tfrac{\hbar^2}{2 m} \bm{\nabla}_{\bm{r}}^2 + \tfrac{1}{2}m\omega^2\bm{r}^2 - q\, 
\bm{r}\cdot\bm{\mathcal{E}}\ .
\end{equation}
To obtain eigenstates/eigenvalues of $H^{(\mathcal{E})}$,
we diagonalize it by means  of the transformation 
$\bm{r}=\bm{r}'+{q \be}/{m \omega ^2}$ resulting in $H^{(\mathcal{E})} (\bm{r}) = H^{(0)} (\bm{r}') - {\alpha\mathcal{E}^2}/{2}$. 
Thus, the eigenstates and eigenvalues of $H^{(\mathcal{E})} (\bm{r})$ are given by $\psi_{\bm{n}}^{_{(\mathcal{E})}}(\br)=\phi_{\bm{n}}^{_{(0)}}(\bm{r}')$ and $\mathcal{U}_{\bm{n}}^{(\mathcal{E})} = \mathcal{U}_{\bm{n}}^{(0)} - {\alpha\mathcal{E}^2}/{2}$, respectively, where $\phi_{\bm{n}}^{_{(0)}}$ and $\mathcal{U}_{\bm{n}}^{(0)}$ denote the eigenstates and eigenvalues of an isolated oscillator~\cite{Atkins_Friedman_book}.
Similar to Eq.~\ref{nret-energy-in-f-x}, the constant energy shift in $\mathcal{U}_{\bm{n}}^{(\mathcal{E})}$ arises due to the static dipole induced by the external electric field. The matrix elements of the dipole moment, $\bm{\mu} = q\,\bm{r}$, are obtained by
\begin{equation}
\label{xmatel}
\hspace{-0.3cm}
\langle l|\mu_x|k\rangle = \,q\,\sigma\left[\textstyle{\sqrt{l}}\,\delta_{l,k+1}\!+\!\textstyle{\sqrt{l+1}}\,\delta_{l,k-1}\right]\!+\! \alpha\,\mathcal{E}_x\,\delta_{k,l}\ ,
\end{equation}
with $\sigma=\sqrt{\hbar/2m\omega}$ as the QDO characteristic length and the bras/kets defined such that $\langle x|i\rangle=\psi_{i}(x)$. In a similar way, one obtains $\mu_y$ and $\mu_z$. The first and second terms on the r.h.s.~of Eq.~\ref{xmatel} correspond to the fluctuating and field-induced static QDO dipoles, respectively.

The Hamiltonian of the total non-interacting system is
\begin{equation}
\label{nonperturbed-H}
H_0=H_{\rm rad}+H_{1}^{(\mathcal{E})}+H_{2}^{(\mathcal{E})} \ ,
\end{equation}
where $H_{\rm rad}$ corresponds to the vacuum radiation field. In the dipole approximation of the multipolar-coupling formalism, the interaction Hamiltonian is given by~\cite{Craig1994}
\begin{equation}
H_{\rm int} = - [ \bm{\mu}_1\cdot\bm{D}_\perp(\bm{r}_1) + \bm{\mu}_2\cdot\bm{D}_\perp(\bm{r}_2)]/\epsilon_0\ .
\end{equation}
Here, $\bd_\perp$ is the transverse component of the vacuum displacement field given by
\begin{equation}
\!\!\!
\bd_\perp(\br) = i \textstyle\sum\limits_{\bk,\lambda}\sqrt{\tfrac{\hbar c k \epsilon_0}{2 V}} \left[\eh_{\bk\lambda} a_{\bk\lambda}^{ }\ex^{i\bk\cdot\br} - \bar{\eh}_{\bk\lambda}
a_{\bk\lambda}^\dagger\ex^{-i\bk\cdot\br}\right]
\end{equation}
with $V$ as the quantization volume. The annihilation and creation operators of a mode with the wave vector
$\bk$ and polarization $\eh_{\bk\lambda}$, $a_{\bk\lambda}$ and $a_{\bk\lambda}^\dagger$, obey the bosonic commutation relations~\cite{Craig1994}. 
The ground-state of $H_0$ is given by
$|0\rangle\!=\!|0,0,0\rangle_1|0,0,0\rangle_2|\{0\}\rangle$,
where the QDO ket states are defined such that  $\langle\bm{r}_i|n_x,n_y,n_z\rangle_i=\psi_{\{n_x,n_y,n_z\}}(\br_i)$, and $|\{0\}\rangle$ is the ground state of the vacuum field. Excited states of the total unperturbed system can be written similarly. By using these states, we follow standard perturbation theory to obtain interaction energies distinguishing between different contributions resulting from coupling of either fluctuating or field-induced static QDO dipole moments, given by Eq.~\ref{xmatel}, to the vacuum field.

The 1st- and 3rd-order corrections vanish since in such cases $a_{\bk\lambda}$ and $a_{\bk\lambda}^\dagger$ occur between two
identical states of the vacuum field.
The 2nd-order perturbation, 
\begin{equation}
\label{2ndpert}
\mathcal{U}^{(2)}=-\textstyle\sum\limits_{_{I\neq 0}}\tfrac{\langle 0|H_{\rm int}|I\rangle \langle I|H_{\rm int}|0\rangle}{E_I-E_0}\ ,
\end{equation}
yields non-vanishing interaction terms only when the vacuum field in $|I\rangle$ is in a single-photon excitation and both QDOs are in their ground states, namely
$|I\rangle=|0,0,0\rangle_1|0,0,0\rangle_2|\bm{1}_{k\lambda}\rangle$.
For this case, after removing the self--energies, $\Delta \mathcal{U}(R)=\mathcal{U}(R)-\mathcal{U}(\infty)$, we obtain an interaction energy between the two QDOs given by
\begin{align}
\label{ES-QED}
\Delta \mathcal{U}^{(2)}(R)=\tfrac{\alpha_1\alpha_2(\mathcal{E}_x^2+\mathcal{E}_y^2-2 \mathcal{E}_z^2)}{4\pi\epsilon_0 R^3}\ ,
\end{align}
which is the same as the second term of Eq.~\ref{nret-energy}.
Considering the QDO states in $|I\rangle$, this interaction energy corresponds to the situation when both QDOs couple to the vacuum field via their static field-induced dipoles and exchange one virtual photon, indicating the electrostatic nature of this interaction term. Taking into account the $R^{-3}$ scaling, this term corresponds to a field-induced (dipole-dipole) electrostatic interaction.

For the 4th-order correction we have two terms~\cite{Craig1994}
\begin{align}
\label{4th-pert}
\hspace{-0.225cm} \mathcal{U}^{(4)}\!= -\hspace{-0.2cm}\textstyle\sum\limits_{{I,I\!\!I,I\!\!I\!\!I}\neq 0}\hspace{-0.3cm}
\tfrac{
	\langle 0|H_{\rm int}|I\!\!I\!\!I\rangle 
	\langle I\!\!I\!\!I|H_{\rm int}|I\!\!I\rangle
	\langle I\!\!I|H_{\rm int}|I\rangle 
	\langle I|H_{\rm int}|0\rangle
}{(E_I-E_0)(E_{I\!\!I}-E_0)(E_{I\!\!I\!\!I}-E_0)}\ \ +
\textstyle\sum\limits_{{I,I\!\!I}\neq 0}
\tfrac{
	\langle 0|H_{\rm int}|I\!\!I\rangle 
	\langle I\!\!I|H_{\rm int}|0\rangle
	\langle 0|H_{\rm int}|I\rangle 
	\langle I|H_{\rm int}|0\rangle
}{(E_I-E_0)^2(E_{I\!\!I}-E_0)}\ .
\hspace{-0.2cm}
\end{align}
For non-polar species, only the first term of Eq.~\ref{4th-pert} is relevant and hence widely used in the literature. The second term becomes relevant for polar atoms/molecules with permanent electrostatic moments, as they can couple to the vacuum field via either fluctuating or static dipoles~\cite{Craig1994, Passante2018}. In the latter case, the interacting species can emit/absorb a virtual photon without undergoing a change in their energy eigenstate.
In our case, the two QDOs possess static field-induced dipoles due to the applied electric field and the second term in Eq.~\ref{4th-pert} plays an important role. Specifically, when each QDO couples to the vacuum field via its static field-induced dipole, both 4th-order terms
in Eq.~\ref{ES-QED} yield non-vanishing contributions $\propto R^{-5}$ of the same magnitude but opposite sign, cancelling each other~\cite{Our_arXiv}. 
When both atoms couple to the vacuum field via their fluctuating dipoles, the treatment of Eq.~\ref{4th-pert} as in Refs.~\cite{Salam2009,Craig1994} delivers~\cite{Our_arXiv} the known London and Casimir-Polder dispersion energies
\begin{align}
\label{dispersion-qed}
\Delta \mathcal{U}_{1}^{(4)}(R) = 
\left\{ \begin{matrix} -\frac{3\alpha_1 \alpha_2 ~ \omega_1 \omega_2 \hbar }{2\, (4\pi\epsilon_0)^2\, (\omega_1+\omega_2) R^6}~
\ ,\ \ R \ll c/\omega \ , \\
-\frac{23 \hbar c ~\alpha_1\alpha_2}{4\pi\, (4\pi\epsilon_0)^2 R^7}\qquad\ \ \ ,\ \ \ R \gg c/\omega\ ,\end{matrix}\right.
\end{align}
respectively. The former is the first term of Eq.~\ref{nret-energy} and the latter is its counterpart for the 
retarded regime, where the frequencies disappear since at large distances each species effectively ``sees'' another one as a static object. 
Finally, when one of the species couples to the vacuum field via its static dipole moment and the other one by its fluctuating dipole, the resulting interaction energy is
\begin{equation}
\label{induction-qed}
\Delta \mathcal{U}_{2}^{(4)}(R)=
-\tfrac{\alpha_1 \alpha_2(\alpha_1+\alpha_2)~(\mathcal{E}_x^2+\mathcal{E}_y^2+4\mathcal{E}_z^2)}{2 [4\pi\epsilon_0]^2 R^6}\ ,
\end{equation}
which is the third term of Eq.~\ref{nret-energy}. 
This interaction stems from the 4th-order correction and has $R^{-6}$ scaling, similar to the nonretarded dispersion energy, but in contrast to the latter remains unaffected by retardation. These features allow us to identify the term of Eq.~\ref{induction-qed} as the field-induced polarization energy. 
Within the employed QED approach, one can also take into account the influence of a static field on the cloud of virtual photons surrounding each atom~\cite{Passante1985,Compagno1985,Passante1987,Compagno1988,Compagno1992}. Although the effect of distortions of such photon clouds on the molecular interactions  
is usually assumed to be small, it might become important under external fields, similar to the aforementioned effects caused by hyperpolarizabilities. Moreover, the impact of clouds of virtual photons on the molecular interactions can become even more important for atoms at close separation, \textit{i.e.} when electron exchange effects become crucial. Indeed, the distortion of photon clouds by a static field can change the effective atomic vdW radii, but to investigate such effects in detail~\cite{Tkatchenko2021} one would need to go beyond the perturbative QED approach used here.

In addition to the two quantum approaches discussed above, we present a (semi)classical derivation that allows us to transparently connect different molecular interactions to the fields originating them.
To this end, we employ a SED approach developed by Boyer~\cite{Boyer1969,Boyer1971} based on the theory of classical electrodynamics with a random zero-point radiation field. Within this picture, the fluctuating vacuum field induces random polarization of atoms (modeled by classical oscillators) and couples them to each other through their electromagnetic fields, as described in classical electrodynamics. For the retarded regime, $R \gg c/\omega$, only large wavelengths contribute to
the interatomic interaction and the SED equation of motion for a dipole oscillator reduces to $m \omega^2 \bm{r} = q \bm{E}(\bm{r},t)$.
Solving this equation yields $\bm{\mu}(\bm{r},t)=\alpha\bm{E}(\bm{r},t)$, with $\bm{E}$
as the total electric field at point $\bm{r}$ and time $t$.
The energy of the electric dipole moment induced by $\bm{E}$ in the same 
field is known from electrodynamics as $\mathcal{U}=-\frac{1}{2}\alpha\langle\bm{E}^2\rangle$, where the bracket indicates time averaging.

Since we apply a static electric field on top of the random radiation field, the oscillator dipole has two parts, each related to one of these fields. According to Fig.~\ref{fig:QDOs}, the first QDO is
located at the origin, while we bring the second QDO to the point $\bm{r}_2=(0,0,R)$ from $z = + \infty$. The difference
in the energy of the oscillators for the two configurations, $\Delta\mathcal{U}(R)=\mathcal{U}(R)-\mathcal{U}(\infty)$, is the interaction energy. The total electric field at $\bm{r}_2$ is given by
\begin{equation}
\label{total-E-r2}
\bm{E}(\bm{r}_2,t)= \bm{E}_0(\bm{r}_2,t)+\bm{E}_{\mu_1}(\bm{r}_2,t)+\bm{\mathcal{E}}_{\mu_1}(\bm{r}_2)+\bm{\mathcal{E}}\ ,
\end{equation}
with $\bm{E}$ and $\bm{\mathcal{E}}$ as radiation and static fields, respectively. Here, the random radiation field
is defined by~\cite{Boyer1975}
\begin{align}
\label{zero-point-field}
\bm{E}_0 = \mathrm{Re}\textstyle\sum\limits_{\lambda=1}^{2}\int d^3k \tfrac{\bm{\epsilon}(\bm{k},\lambda)
\ \mathfrak{h}(\bm{k},\lambda)}{\sqrt{4\pi\epsilon_0}}\ \ex^{i[\bm{k}\cdot\bm{r}- k c t+\theta(\bm{k},\lambda)]}\ ,
\end{align}
where, $\mathfrak{h}^2(\bm{k},\lambda)=\hbar k c/2\pi^2$ is the energy of each mode of the random field and the sum runs over the two possible polarizations. The random phase $\theta$ ranges from $0$ to $2\pi$ and $\bm{\epsilon}(\bm{k},\lambda)$ are orthogonal
unit polarization vectors, $\bm{\epsilon}(\bm{k},\lambda)\cdot\bm{\epsilon}(\bm{k}',\lambda')=\delta_{\lambda\lambda'}$.
Then, $\bm{E}_{\mu_1}(\bm{r}_2,t)$ is a time-dependent field radiated by the oscillating dipole of the first species induced by the random radiation field. Similarly, $\bm{\mathcal{E}}_{\mu_1}(\bm{r}_2)$ is the electric field of the static dipole of the first oscillator, which is induced by the uniform external electric field. The electric field of an oscillating electric dipole is~\cite{Jackson1998}
\begin{align}
\begin{array}{ll}
\!\!
\bm{E}_{\mu} (\bm{r},t) = \mathrm{Re}\left[\left(\tfrac{k^2(\bm{n}\times\bm{\mu})\times\bm{n}}{4\pi\epsilon_0 r}
+ (1 - i k r) \bm{\mathcal{E}}_{\mu} \right)\ex^{ikr}\right] ,\!\!
\end{array}
\label{E-p-field}
\end{align}
where $\bm{\mathcal{E}}_{\mu} (\bm{r}) = \left(3\bm{n}(\bm{\mu}\cdot\bm{n})-\bm{\mu}\right)/4\pi\epsilon_0 r^3$ is the electric field of a static electric dipole.
Therefore, the energy of the second oscillator, with the static polarizability $\alpha_2$, under the total electric field given by Eq.~\ref{total-E-r2} is obtained as $\mathcal{U}(\bm{r}_2)= -\frac{1}{2}\alpha_2\left\langle\bm{E}^2(\bm{r}_2,t)\right\rangle$. Then, subtracting energies for $z_2=R$ and $z_2 = +\infty$, yields
\begin{align}
\label{ret-deltaE-t}
\begin{array}{ll}
\Delta\mathcal{U}(R)=-\alpha_2 \left[ \big\langle\bm{E}_0(\bm{R},t)\cdot\bm{E}_{\mu_1}(\bm{R},t)\big\rangle + \big\langle\bm{\mathcal{E}}\cdot\bm{\mathcal{E}}_{\mu_1}(\bm{R})\big\rangle
+\frac{1}{2}\big\langle\bm{\mathcal{E}}_{\mu_1}(\bm{R})\cdot\bm{\mathcal{E}}_{\mu_1}(\bm{R})\big\rangle \right]\ ,
\end{array}
\end{align}
where we keep only nonvanishing terms after time and phase averaging~\cite{Our_arXiv}.
The first term in Eq.~\ref{ret-deltaE-t} gives the interaction energy from the coupling of two fluctuating dipole moments 
induced by the random field. This point was addressed by Boyer~\cite{Boyer1975} who obtained the aforementioned London~\cite{Boyer1972,Boyer2018} and Casimir-Polder~\cite{Boyer1969,Boyer1971,Boyer1973}
results for the nonretarded and retarded regimes, respectively.
Here, we focus on other contributions to the interaction energy. The second term in Eq.~\ref{ret-deltaE-t} is the coupling of the field-induced dipole of the second oscillator with the static field of the field-induced dipole of the first oscillator which gives the electrostatic interaction energy 
\begin{equation}
\label{Energy-E(p1)-E(xt)}
\!\!\!\Delta\mathcal{U}_2(R)\!=\!-\alpha_2\bm{\mathcal{E}}\!\cdot\!\left[\tfrac{3\hat{z}(\alpha_1\bm{\mathcal{E}}\cdot\hat{z})
-\alpha_1\bm{\mathcal{E}}}{4\pi\epsilon_0 R^3}\right]
=\tfrac{\alpha_1\alpha_2(\mathcal{E}_x^2+\mathcal{E}_y^2-2\mathcal{E}_z^2)}{4\pi\epsilon_0 R^3}\ .
\end{equation}
The third term of Eq.~\ref{ret-deltaE-t} describes interaction of the dipole moment of the second oscillator induced by the static field of the first oscillator, with the same field and yields the interaction energy 
\begin{equation}
\begin{aligned}
\label{Energy-E(p1)-E(p1)}
\!\!\!\Delta\mathcal{U}_3(R)\!=\!-\tfrac{\alpha_2}{2}\!\left(\tfrac{3\hat{z}(\alpha_1\bm{\mathcal{E}}\cdot\hat{z})-\alpha_1\bm{\mathcal{E}}}{4\pi\epsilon_0 R^3}\right)^2 \!\!\!=\!
-\tfrac{\alpha_2\alpha_1^2(\mathcal{E}_x^2+\mathcal{E}_y^2+4\mathcal{E}_z^2)}{2(4\pi\epsilon_0)^2R^6}\ .
\end{aligned}
\end{equation}
The mechanism responsible for this interaction is similar to the one for intermolecular  polarization (induction) interaction between molecules with permanent dipole moments, with the difference that here the static dipoles are induced by the external field and hence the interaction is a field-induced polarization interaction. Adding the term coming from the interaction of the dipole moment of the first oscillator with the static field of the field-induced dipole of the second oscillator, yields the full field-induced polarization energy given by Eq.~\ref{induction-qed}.
The polarization energy of Eq.~\ref{Energy-E(p1)-E(p1)} was originally derived from a pure classical point of view, by considering the coupling of a \emph{pair} of polarizable objects, interacting and therefore possessing the collisional contribution $\Delta \alpha$ to the \emph{pair polarizability}, to an external static field~\cite{Silberstein_1917_1, Silberstein_1917_2, Buckingham_1956, Buckingham_1973, Buckingham_1978, Hunt1985, Hunt1986}.

\begin{figure*}[t!]
\includegraphics[width=0.48\linewidth]{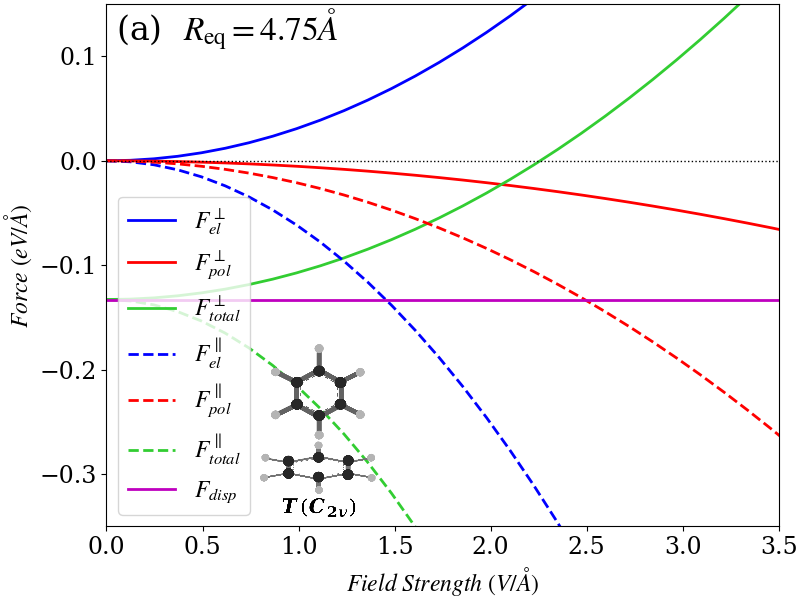}
\quad
\includegraphics[width=0.48\linewidth]{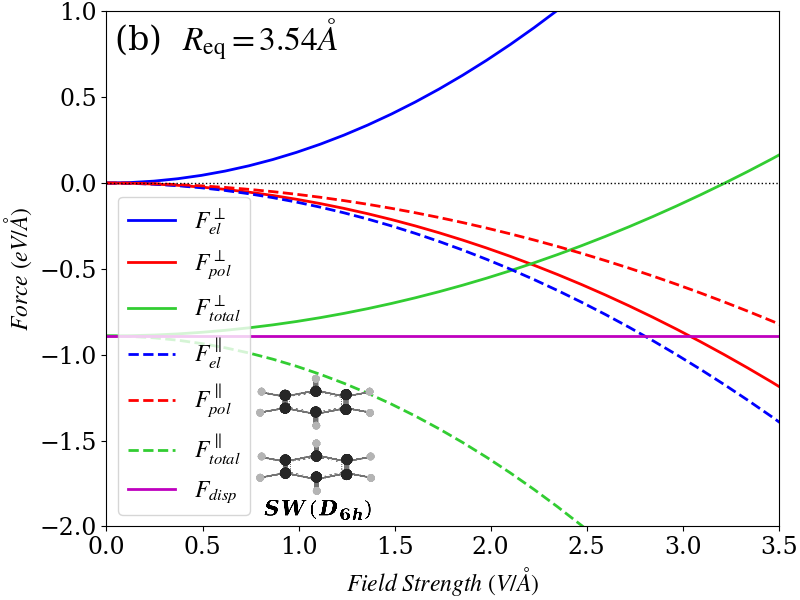}
\caption{The dispersion force
($F_{\rm disp}$) as well as the field-induced electrostatic ($F_{\rm el}$) and polarization ($F_{\rm pol}$) forces for two interacting benzene molecules separated by the corresponding equilibrium distance 
$R_{\rm eq}=4.75$\AA\ and $R_{\rm eq}=3.54$\AA\ within {\bf (a)} T-Shape and {\bf (b)} Sandwich structure, respectively. The results are shown for two different alignments of the external field and the molecules: the field is either parallel $(||)$ or perpendicular $(\perp)$ to the line connecting the molecule centers.}
\label{fig:Benzene}
\end{figure*}

Our results, obtained within the QDO model for the linear-response regime, show that the field-induced electrostatic and polarization energies are not influenced by the retardation effects and only the conventional leading-order London/Casimir dispersion energy changes the distance scaling law going from the nonretarded to retarded regime.
However, without aforementioned contributions coming from dispersion--induced changes in the polarizability~\cite{Jansen1955, Buckingham_1956, Buckingham_1973, Buckingham_1978, Certain1971, Hunt1980, Hunt1981, Hunt1986, Fowler1994, Hunt_1994, Joslin_1996} and dipole moments~\cite{Feynman1939, Hunt_1980_dipole, Galatry1980, Galatry1983, Hunt1985, Craig_1981, Hunt_1987, Hunt_1987_2, Fowler_1990, Vigoureux1983, Odbadrakh2016} of interacting species as well as field-induced  hyperpolarization effects~\cite{Salam1997, Hu2021}, the dispersion interaction remains unchanged by the external uniform static field, for both regimes. 
This fact is in contradiction to the  recent results of Fiscelli \emph{et al.}~\cite{Fiscelli2020}, who obtained a drastic change in the distance dependence of dispersion interaction energy between two atoms after applying a static electric field ($R^{-3}$/$R^{-4}$ for the nonretarded/retarded regime). 
In Ref.~\citenum{Fiscelli2020}, the wavefunctions of a two-level \lq\lq hydrogen\rq\rq\ atom in a static electric field were obtained from perturbation theory. 
The interaction energy between two atoms in a field was calculated from QED perturbation theory using the atomic wavefunctions obtained in the first step.~However, since these functions do not form a complete set,
employing them as a basis set for the second use of the perturbation theory in Ref.~\citenum{Fiscelli2020} is questionable. A similar conclusion was recently obtained in Ref.~\citenum{Hu2021} independently of our work~\cite{Our_arXiv}.

Due to the analytical solutions of the QDOs, the results of Eq.~\ref{nret-energy} can be straightforwardly generalized to any number of QDOs, with each of them under their own static field. This provides an opportunity to effectively model internal atom-dependent electric fields present in large molecules. Moreover, the developed framework paves the way for tuning intermolecular interactions by means of external static electric fields, which is important for many practical applications including biophysics~\cite{JCP-StatPhys2016}. This is demonstrated in Fig.~\ref{fig:Benzene}, where the intermolecular forces in a benzene dimer are considered for its two different equilibrium structures, the so called \lq\lq T-Shape\rq\rq\ and \lq\lq Sandwich\rq\rq~\cite{benzene1,benzene2}. The interplay between the field-induced forces and the dispersion force can lead either to cooperation or competition. 
In the case of the \lq\lq T-Shape\rq\rq\ structure, the total electrostatic force can overtake the dispersion force, if a static electric field with strength 2.25\,V/\AA\ is applied perpendicularly to the dimer. 
For the \lq\lq Sandwich\rq\rq\ configuration, the strength of the perpendicular compensating field is around 3.22\,V/\AA.
Although for the considered molecular dimers at the equilibrium distance these fields are relatively strong, 
they are still weaker than internal molecular fields experienced by valence electrons.
Moreover, as shown in Ref.~\citenum{Our_arXiv}, the field required for mutual compensation of the intermolecular forces becomes much weaker at larger separations. Furthermore, for large molecules or nanoscale systems with a multiplicity
of normal modes contributing to Eq.~\ref{nret-energy-in-f-x}, 
a significantly smaller magnitude of the electric field will be required to overcome the dispersion attraction~\cite{Ambrosetti-Science}. Altogether, this shows that applied static electric fields can be used to influence the stability and dynamics of complex molecular system.

Before summarizing our results, it is important to emphasize some of the remaining limitations of the employed coupled QDO model, whose resolution would be needed to form a complete physical picture of intermolecular interactions under a static electric field. As was exhaustively discussed in Ref.~\citenum{Jones2013}, a QDO does possess multipole hyperpolarizabilities starting from the dipole-dipole-quadrupole one, but the first ($\beta$) and second ($\gamma$) hyperpolarizabilities vanish due to the spherical symmetry and Gaussian wavefunction, respectively. Therefore, a single QDO does not fully capture the contributions to the intermolecular interaction energy which are related to hyperpolarization effects of interacting atoms and molecules in static fields. 
For a pair of two-level hydrogen atoms it is known~\cite{Hu2021} that the energy contribution from the field-induced $\beta$ hyperpolarizabilities of the atoms scales as $R^{-11}$. Therefore, this contribution can be neglected in comparison to the dispersion and field-induced electrostatic and polarization interactions. It has been shown that dispersion interactions modify the polarizability of an interacting pair where the leading-order correction scales as $R^{-6}$ and depends on the $\gamma$ hyperpolarizability of atoms or molecules \cite{Jansen1955, Buckingham_1956, Buckingham_1973, Buckingham_1978, Certain1971, Hunt1980, Hunt1981, Hunt1986, Fowler1994, Hunt_1994, Joslin_1996}. In a static electric field, the dispersion-induced polarizability yields an additional interaction energy (quadratic in the applied field) that scales as $R^{-6}$ being comparable~\cite{Buckingham_1978, Joslin_1996} to the field-induced polarization energy in Eq.~\ref{nret-energy}.
To capture a complete picture of the dipolar interactions in a static field, one could use more than one QDO to represent atoms or molecules,
which brings anharmonicity to the system and breaks the spherical symmetry of the model when interacting with external fields.
Such a multi-QDO model would exhibit $\beta$ and $\gamma$ hyperpolarizabilities and thus provide more realistic response properties in comparison to single QDOs. The dipolar interactions between two (or many) multi-QDO systems can be expressed as a coupled-QDO problem, which is still exactly solvable. 

In addition, the considered contributions to molecular interactions between atoms or molecules represented by QDOs were derived assuming large interspecies distances, effectively neglecting exchange and overlap effects. Within the QDO model, an inclusion of the exchange interaction is possible with a further generalization of the existing formalism valid for two identical QDOs~\cite{Fedorov2018,Vaccarelli2021} to the case of heteronuclear species. The overlap effects can be also included following the work of Refs.~\citenum{Hunt1980,Hunt1984}.

In summary, we presented dominant contributions to the interaction energy between two closed-shell atoms or molecules in a uniform static electric field for both, short (nonretarded regime) and long (retarded regime) separation distances. 
Based on first principles of quantum mechanics and QED, and employing the QDO model as a reliable tool for modeling atomic/molecular responses, our framework admits many generalizations. In addition to external fields, the internal field from atomic charges and dipoles within a molecule can be treated as well, leading to an efficient many-body model of different types of intramolecular and intermolecular interactions on equal footing. More general time-dependent fields can also be included. A particularly interesting and novel research direction is to develop and apply a QED treatment to many-body states for a set of many interacting QDOs. This could generalize the transition from the London to Casimir regime for an assembly of atoms or molecules beyond the well-known two-body case. 
All in all, we are confident that our framework barely scratches the surface of possible developments and applications in the field of molecular interactions under the combined action of external and vacuum fields.

\begin{acknowledgement}
The authors acknowledge the financial support from the Luxembourg National Research Fund through the FNR CORE projects \lq\lq QUANTION(C16/MS/11360857, GrNum:11360857)\rq\rq\ and
\lq\lq PINTA(C17/MS/11686718)\rq\rq\ as well as from the European Research Council via ERC Consolidator Grant \lq\lq BeStMo(GA n725291)\rq\rq .
\end{acknowledgement}

%
%
%

\bibliography{Literature_JPCL_QDO}

\end{document}